\documentclass[
   ,final            
  ]
  {aipproc}

\layoutstyle{6x9}

\begin{document}

\title{The Yale LAr TPC}


\author{A. Curioni}{
  address={Yale University}
}

\author{B. T. Fleming}{
  address={Yale University}
}

\author{M. Soderberg}{
  address={Yale University}
}

\begin{abstract}
In this paper we give a concise description of a liquid argon
time projection chamber (LAr TPC) developed at Yale, and present
results from its first calibration run with cosmic rays.
\end{abstract}

\maketitle


\section{Introduction}

Liquid argon time projection chambers (LAr TPC) are nearly optimal
detectors for neutrino experiments looking for $\nu _e$ appearance on a
$\nu_{\mu}$ beam in the energy range 0.5 -- 5 GeV. The LAr TPC
technology has been proposed for the measurement of $\theta _{13}$, CP
violation in the neutrino sector and  determination of the mass
hierarchy (e.g.~\cite{proposal1, proposal2, proposal3}), and to study
the MiniBooNE low energy anomaly \cite{miniboone, microboone}.  
The technique is equally promising for proton decay searches
\cite{bueno}.   
A LAr TPC for neutrino physics was first proposed in 1977 by Carlo
Rubbia and a vigorous R\&D program was then established, which
produced decisive steps in defining the technology and its
applicability to particle physics (e.g.~\cite{rubbia, 3ton, 50l,
  t600}).   
Very remarkably, images taken in a LAr TPC are comparable in quality 
with pictures from bubble chambers. As for bubble chambers, events can
be analyzed reconstructing 3-momentum and particle type for each track
in the event image, down to low energy (few MeV for electrons,
few tens of MeV for protons); the calorimetric performance ranges from
good to excellent, depending on event energy and topology.   
Equally important, the LAr TPC technology is suitable for very massive  
(several 10 ktons) detectors, as required in most contemporary
neutrino physics. 
A very active R\&D program is on going in the U.S.; as part of this
larger effort, a LAr TPC has been developed at Yale starting in
2005. The detector has been commissioned in early 2007, and cosmic ray
tracks have been imaged. Preliminary results are presented here.  

\section{Description of the detector}

\begin{figure}[htb]

  \includegraphics[width=.32\textwidth]{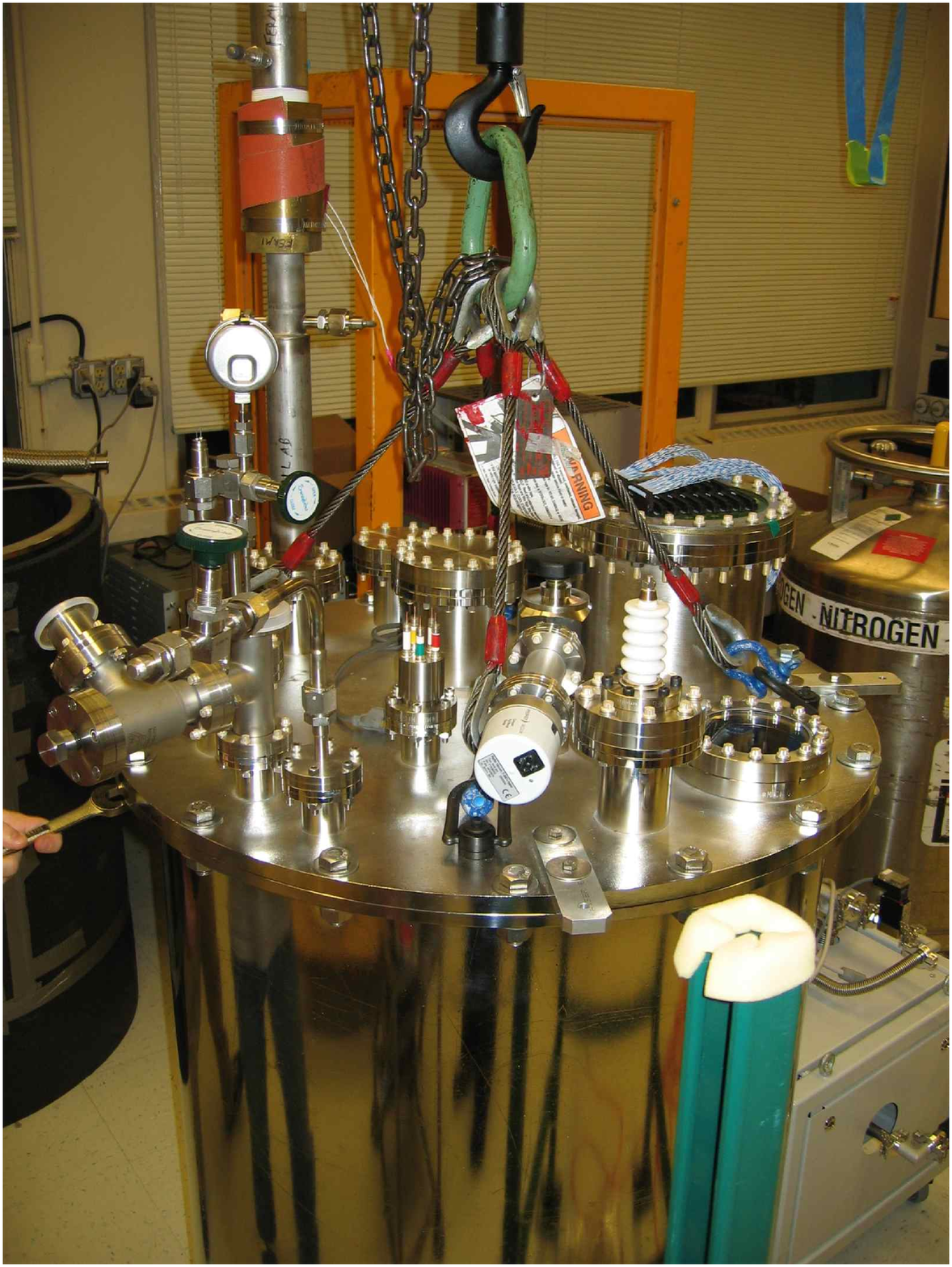}
~~~~~~~   \includegraphics[width=.55\textwidth]{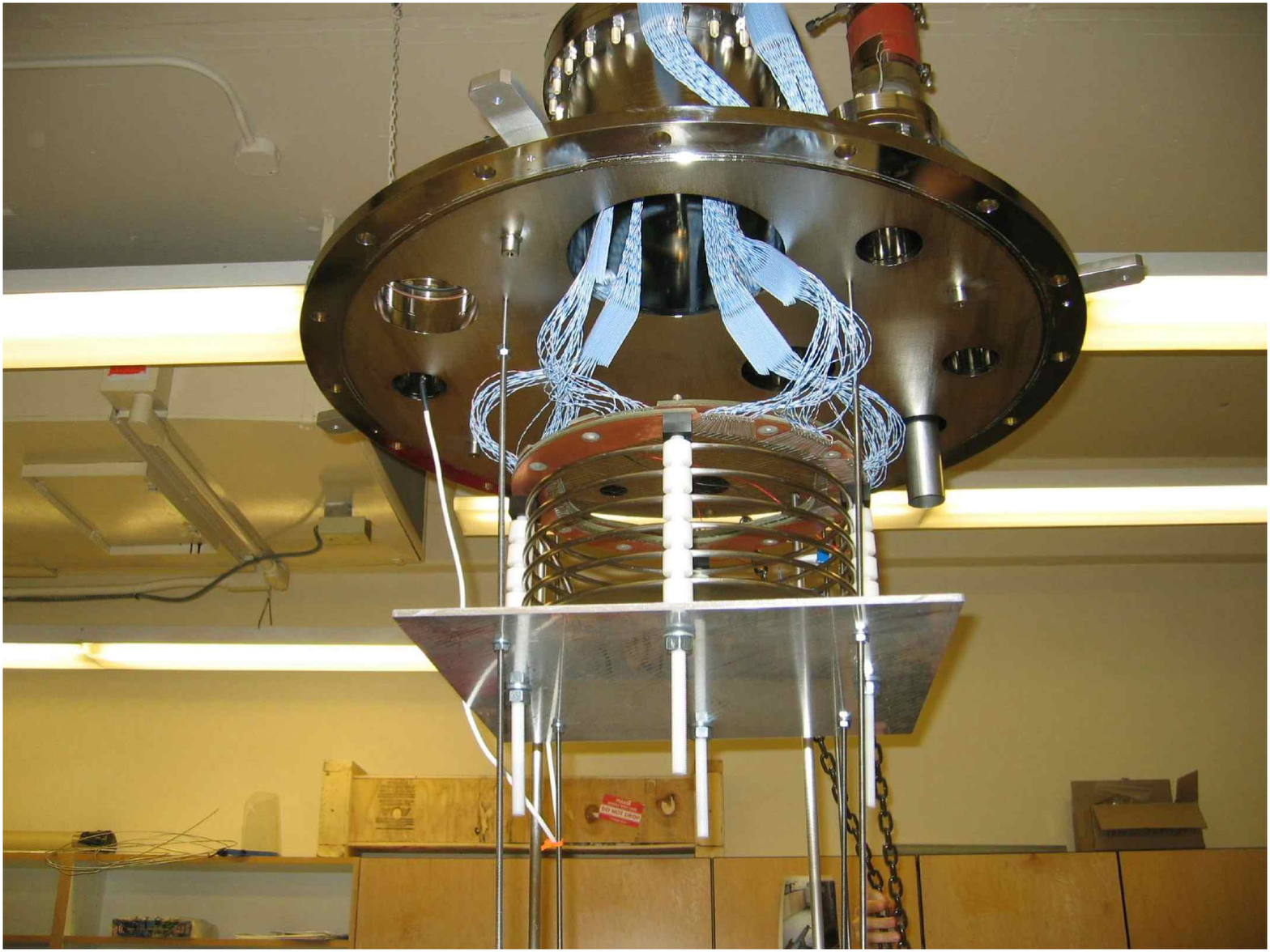}
  \caption{{\it Left:} the top flange, described in the text; {\it
  Right:} the TPC, fully cabled, hanging from the top flange. }
\label{f1}
\end{figure}


\begin{enumerate}

\item
{\it Cryogenics:} 
the LAr TPC is housed in a cylindrical stainless steel vessel of total
volume 500~l. The inside walls are electropolished. The vessel is
evacuated to few 10$^{-6}$ mbar (at LAr temperature, 87 K) prior to
filling with ultra pure LAr. The vessel is cooled to LAr temperature
by an open bath filled with commercial, non-purified LAr.  
It takes about five hours to cool down the system and fill it until
the TPC is fully covered (about 250~l of ultra-pure LAr). The total
LAr consumption is $\sim$ 1,000~l for a 24 hr long experiment.  
The top flange of the TPC vessel (Fig.~\ref{f1}-$left$) houses several
ports: 
feedthroughs for the TPC high-voltage, signal cables, test pulse and
capacitive level meter, high voltage feedthrough and optical
feedthrough for a purity monitor mounted inside the vessel, pumping
line, filling line, relief valve, pressure gauges, and a window. All
the seals are CF or VCR. The top flange itself is sealed using a Viton
O-ring, therefore the system is not designed to be vacuum-tight at LAr 
temperature in steady state. The adopted filling procedure is to break
the vacuum using ultra-pure cold Ar gas while the LAr open bath is
about half full. In steady state the system runs at an overpressure of
0.3 atm. The relief valve has been provided by H. Jostlein of FNAL.  
%

\item
{\it Liquid Ar purification:}
the necessity to drift free electrons in LAr for {\it O}(1
millisecond) or longer sets very stringent requirements on the purity
of LAr, at the level of several tens of parts per trillion of
O$_2$-equivalent contamination (cf. few parts per million in
commercially available LAr).     
The filters to purify LAr have been developed and built at FNAL and 
are described in \cite{pordes}; a detailed paper is in
preparation. The filter itself is made of a copper alumina catalyst
\footnote{Engelhard Copper Alumina catalyst CU-0226S}, packaged in  a
CF nipple with sinterized steel caps. The filter is regenerated in
place at Yale. Commercial grade LAr is purified with a single pass
through the filter, without any additional molecular sieve, at a rate
of $\sim$60~l/hr.   
To measure the electron lifetime independently of the TPC operation,
a purity monitor (see \cite{t600}, and provided by the FNAL group) was
mounted underneath the TPC.   
A loss of drifting charge due to attachment to impurities of less than
10\% over a drift time of 0.5~ms has been repeatedly measured in the
TPC vessel, stable over a period of 24 hours without recirculation.  

\item
{\it TPC and Electronics:}
the fiducial volume of the LAr TPC (Fig.~\ref{f1}-$right$) is a
cylinder of 33~cm diameter and 17~cm tall. The field cage is made of 6
hollow stainless steel rings, separated by Teflon spacers, and
connected through a chain of 100~M$\Omega$ resistors. 
There are two parallel readout wire planes, with 50 wires each and a
wire pitch of 5 mm.   
Flat cables are soldered to the wire planes and reach the readout
electronics through a signal feedthrough (INFN-Padova / ICARUS),
which holds 512 channels.
The readout electronics is the same as the ICARUS one and has been
provided by the INFN-Padova / ICARUS group \cite{t600}.  
%

\end{enumerate}


\section{Results}

\begin{figure}[htb]
  \includegraphics[width=.5\textwidth]{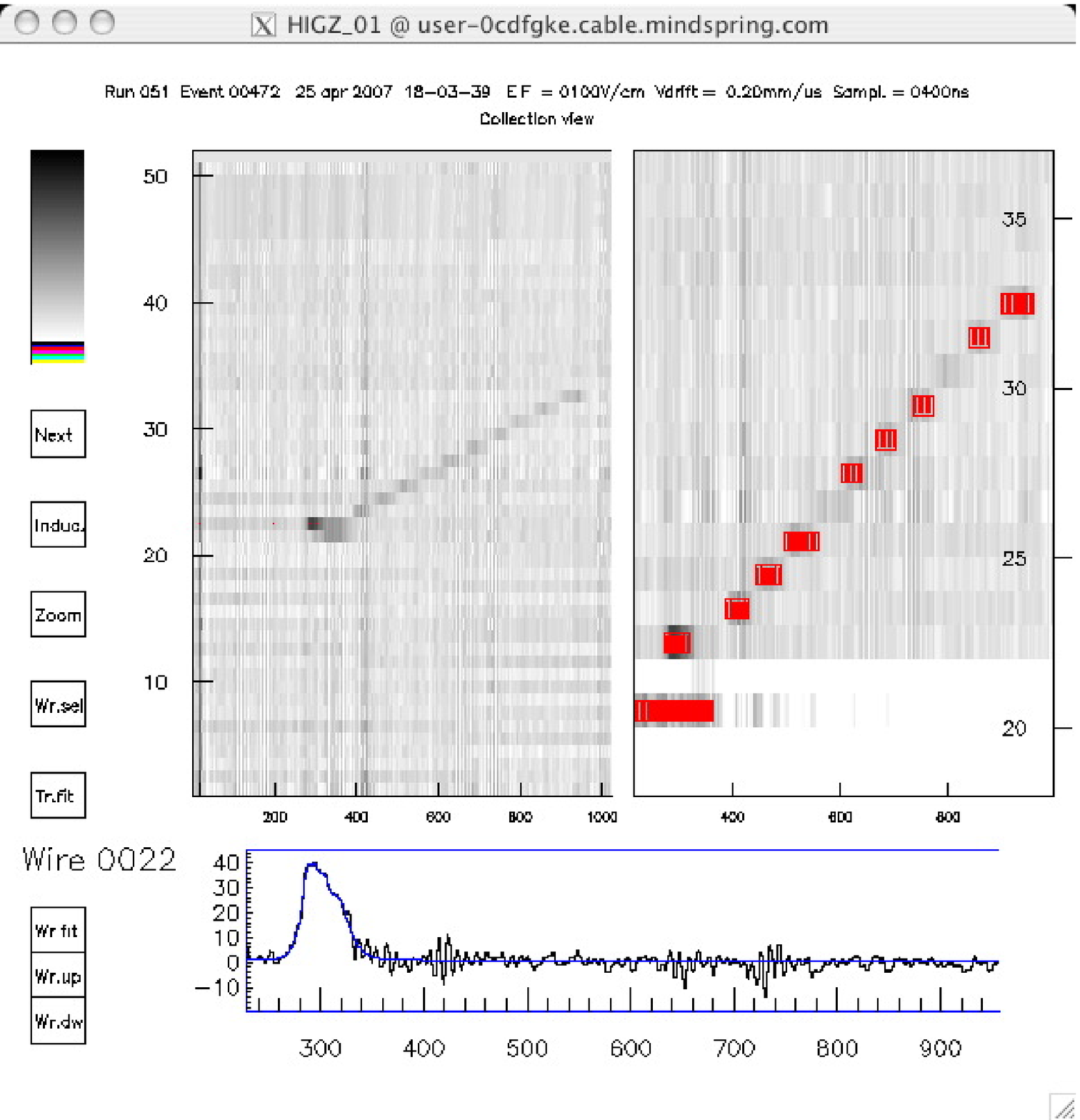}
~~~ \includegraphics[width=.5\textwidth]{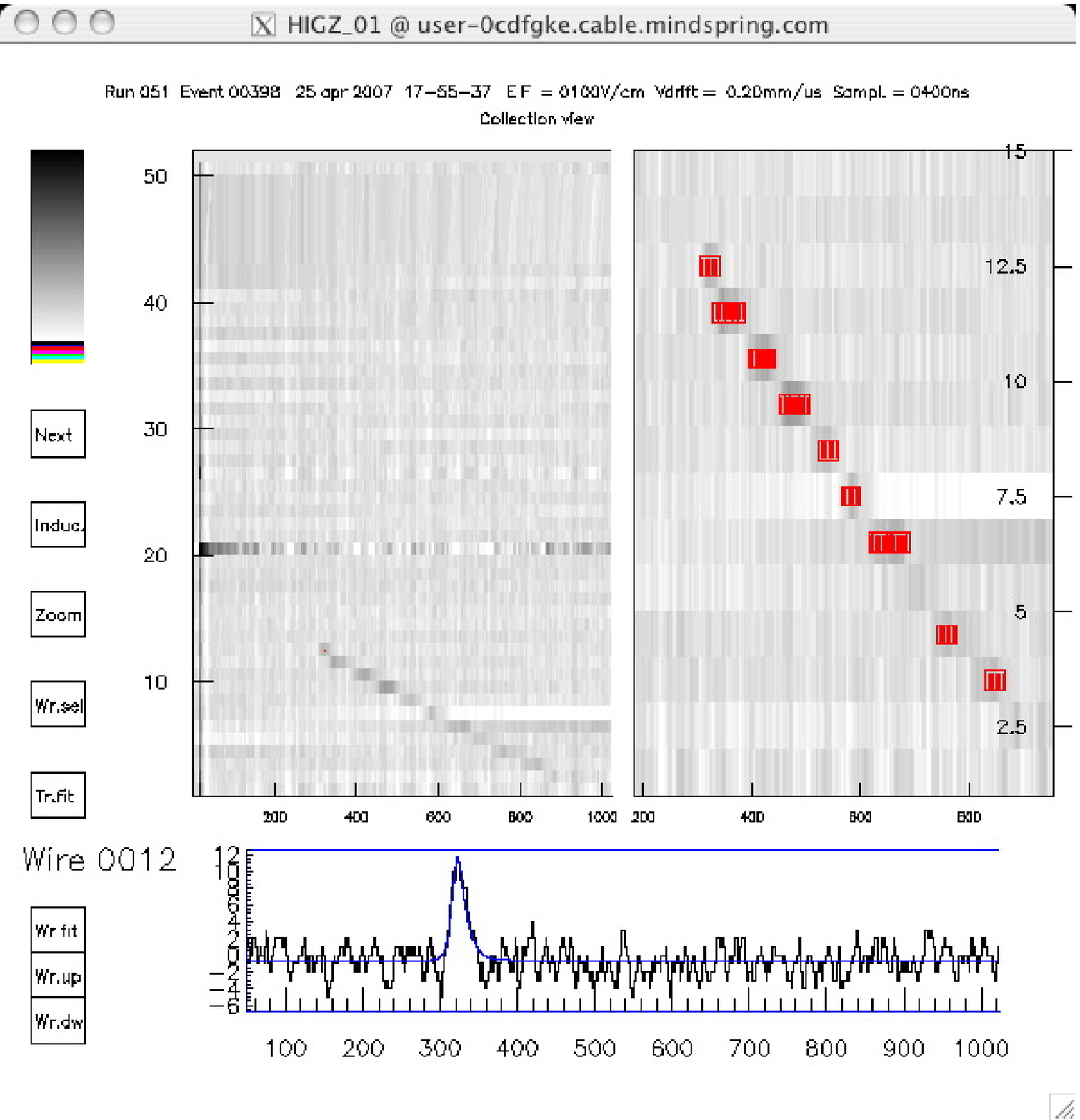}
  \caption{ Two raw images of muons crossing the TPC, displayed as
  wire number {\it vs.} drift time [0.4~$\mu$s]. In both cases the
  zoomed image is also shown; the superimposed {\it squares} show
  the {\it hit position} determined by an automated fit of the
  waveform. The bottom panels show two waveforms, with the fit
  superimposed. }  
\label{f2}
\end{figure}

\begin{figure}
  \includegraphics[width=.82\textwidth]{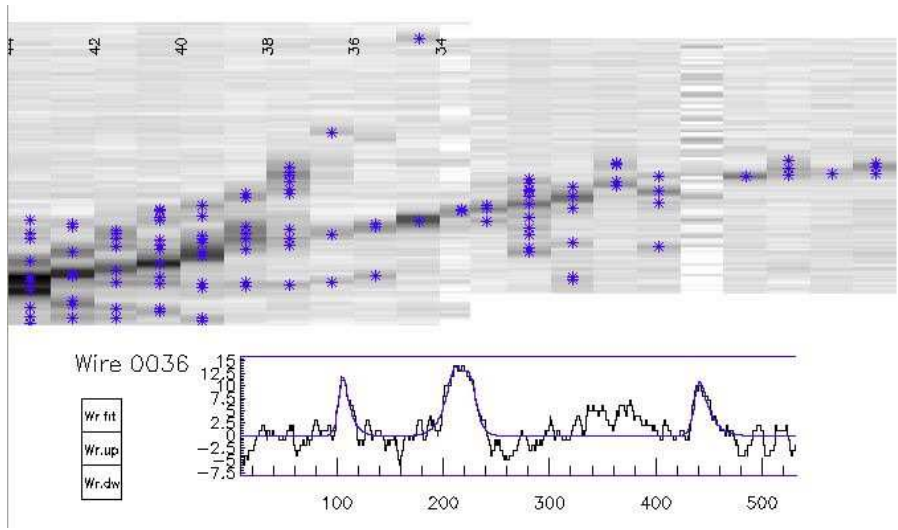}
  \caption{ Raw image of an e.m. shower, displayed as
  drift time [0.4~$\mu$s] {\it vs.} wire number. } 
\label{f3}
\end{figure}

The LAr TPC has been tested on the readily available cosmic rays. 
During data taking the electric field in the drift region was 100~
V/cm, and 250~V/cm between the induction and the collection planes. 
The signals from the induction plane were rather noisy due to
capacitive coupling with the high voltage in the drift region, and are
not presented here. 
The data acquisition was triggered on the sum coincidence of one board
(25 channels) of the collection plane.  
Some events recorded during a cosmic ray run are shown in
Fig.~\ref{f2} and \ref{f3}. The signal is well visible over noise on
each of the hit wires.  
The drift velocity in LAr at 100~V/cm is 0.5~mm/$\mu$s, with a maximum 
drift time of 340~$\mu$s.  

\section{Conclusions}

A prototype LAr TPC has been designed, built and tested at Yale over a
period of two years. It has been developed as a handy R\&D tool for
the university setting, with the possibility of repeated runs over a
short period of time. The LAr purification is a recent development,
applied for the first time to a working imaging instrument, while  the
readout electronics has been provided by the ICARUS collaboration.
The success in imaging cosmic rays marks an important milestone in
terms of technology transfer for the US LAr TPC effort.    
%


\begin{theacknowledgments}
We thank all the people who contributed to this project. In
particular, we acknowledge essential help from: S. Centro, S. Ventura,
B. Baibussinov and the ICARUS group at INFN Padova, for the 
readout electronics, software for DAQ and event display, and
round-the-clock support; N. Canci and F. Arneodo of LNGS, in the
initial work on LAr purification; the FNAL LArTPC group, in particular
H. Jostlein, C. Kendziora and S. Pordes, for providing the filters,
various equipment, and many excellent suggestions; M. Harrison,
S. Cahn and C. E. Anderson of Yale; L. Bartosek of Bartosek
Engineering. 
\end{theacknowledgments}

\end{document}